\documentclass[12pt]{iopart}
\usepackage{caption}
\usepackage{siunitx}
\usepackage{graphicx}
\usepackage{subcaption}

\usepackage{amssymb}
\usepackage{diagbox}
\usepackage{cancel}
\usepackage{bm}

\usepackage{float}
\usepackage{cite}

\newbox{\bigpicturebox}

\begin{document}

\title[Dividing active and passive particles in nonuniform nutrient environments]{Dividing active and passive particles in nonuniform nutrient environments}

\author{Till Welker \& Holger Stark}

\address{Institut für Theoretische Physik, Technische Universität Berlin, Hardenbergstraße 36, 10623 Berlin, Germany}
\ead{welker@tu-berlin.de}
\vspace{10pt}
\begin{indented}
\item[]November 2023
\end{indented}

\begin{abstract} 
To explore the coupling between a growing population of microorganisms such as \emph{E.\ coli} and a nonuniform nutrient distribution, we formulate a minimalistic model. It consists of active Brownian particles that divide and grow at a nutrient-dependent rate following the Monod equation. The nutrient concentration obeys a diffusion equation with a consumption term and a point source. In this setting the heterogeneity in the nutrient distribution can be tuned by the diffusion coefficient. 

In particle-based simulations, we demonstrate that passive and weakly active particles form proliferation-induced clusters when the nutrient is localized, without relying on further mechanisms such as chemotaxis or adhesion. In contrast, strongly active particles disperse in the whole system during their lifetime and no clustering is present. The steady population is unaffected by activity or nonuniform nutrient distribution and only determined by the ratio of nutrient influx and bacterial death. However, the transient dynamics strongly depends on the nutrient distribution and activity. Passive particles in almost uniform nutrient profiles display a strong population overshoot, with clusters forming all over the system. In contrast, when slowly diffusing nutrients remain centred around the source, the bacterial population quickly approaches the steady state due to its strong coupling to the nutrient. Conversely, the population overshoot of highly active particles becomes stronger when the nutrient localisation increases. We successfully map the transient population dynamics onto a uniform model where the effect of the nonuniform nutrient and bacterial distributions are rationalized by two effective areas.
\end{abstract}

\vspace{2pc}

\section{Introduction}
Living microorganisms show two distinct types of activity: motility~\cite{vicsek2012collective, elgeti2015physics, das2020introduction} and proliferation~\cite{hallatschek2023proliferating, BACTERIAL_GROWTH, ranft2010fluidization}. On the one hand, motile microorganisms locally consume chemical energy, use it to self-propel, and dissipate it due to friction with the fluid surrounding. This flux of energy drives the system out of equilibrium. As a consequence, particles with purely repulsive interaction can show motility-induced phase separation~\cite{cates2015motility, buttinoni2013dynamical, fily2012athermal}. Combining motility with mechanical, chemical, hydrodynamic, or social interactions~\cite{zottl2023modeling} results in a rich palette of collective phenomena from turbulent behaviour~\cite{wolgemuth2008collective,ACTIVE_TURBULANCE}, over dynamic clustering~\cite{theurkauff2012dynamic,pohl2014dynamic}, fluid pumps~\cite{nash2010run,hennes2014self}, and convective rolls~\cite{ruhle2020emergent}, to flocking dynamics~\cite{vicsek1995novel,toner1998flocks,knevzevic2022collective, kreienkamp2022clustering}, swirls~\cite{ziepke2022multi,tunstrom2013collective,bittl2023swirling,czirok1996formation}, and fluid clusters caused by non-reciprocal torques~\cite{zhang2021active}. The tendency of chemotactic bacteria to accumulate in regions of high attractant concentration has been known since the 19th century~\cite{engelmann1881neue} and studied ever since with huge interest in self-generated gradients~\cite{marsden2014chemotactic,saragosti2011directional,seyrich2019traveling,sturmer2019chemotaxis}.

On the other hand, proliferation represents a different form of activity. Through cell division and growth, individuals inject biomass into the system, thereby breaking the conservation of mass~\cite{hallatschek2023proliferating}. Mechanical forces due to cell-cell contacts strongly impact the growth process and can be modelled using pressure-dependent growth rates~\cite{shraiman2005mechanical,basan2009homeostatic,ranft2010fluidization} or a cell cycle impacted by pressure~\cite{li2022competition,li2021role}. Cells more resilient to pressure tend to have a competitive advantage in crowded environments~\cite{wagstaff2016mechanical, li2022competition}. 
The \emph{E. coli} bacterium, which we have in mind in our model, tends to be less affected by mechanical pressure compared to a variety of epithelial cells~\cite{li2022competition}, so we neglect pressure in this study. However, it should be noted that mechanical self-regulation slows down the elongation rate~\cite{wittmann2022mechano} and in narrow long channels the mechanical forces become the main limiting factor of cell growth~\cite{yang2018analysis}. Furthermore, for elongated cells the competition between active and passive forces results in a mosaic of locally ordered patches~\cite{you2018geometry}. Finally, proliferation and cell death can also liquidise tissue leading to cell diffusion~\cite{ranft2010fluidization} and the transport of cells is additionally amplified during the transition of the colony from two to three dimensions~\cite{dhar2022self}.

The interplay between motility and proliferation has already shown interesting dynamics and is crucial in describing and interpreting experimental observations. The combination of density-dependent velocity and growth results in clusters of characteristic size and transient ring patterns~\cite{cates2010arrested}, a behaviour previously linked to chemotaxis~\cite{budrene1991complex, woodward1995spatio,budrene1995dynamics}. Furthermore, a system of reaction-diffusion equations modeling growing bacterial colonies in the presence of nutrient~\cite{kawasaki1997modeling} demonstrates that motility is crucial to describe the rich variety of spatial patterns observed in experiments~\cite{wakita1994experimental, ohgiwari1992morphological,kawasaki1997modeling}. Whereas, immobilizing bacteria reduces variety of possible colony patterns~\cite{wakita1994experimental}. 

As soon as cell number is no longer conserved, another aspect becomes relevant - population dynamics. Classic approaches to model it range in complexity from exponential growth for abundant resources, over density-dependent growth rates~\cite{BACTERIAL_GROWTH} and coupled differential equations for population-nutrient~\cite{CHEMOSTATEN_THEORIE} or predator-prey dynamics~\cite{volterra1926fluctuations}, to models taking into account stochastic fluctuations~\cite{khatri2012oscillating} or~spatial structure~\cite{fisher1937wave}. The latter has significant impact on population dynamics of fish populations~\cite{goethel2011incorporating}, host-parasitoid systems~\cite{hassell1991spatial}, and on the spread of infectious diseases~\cite{grenfell1995spatial}. A recent study combined motility and proliferation to model the transition from initial exponential to subexponential growth~\cite{schnyder2020control}.

In this article, we study the population dynamics of active agents and their emergent collective patterns in a non-uniform nutrient profile using active Brownian particles. Hydrodynamic interactions due to shape-dependent flow fields~\cite{ELONGATED_SWIMMERS,COLLECTIVE_ELONGATED} are neglected. To explore purely proliferation-induced effects, we ignore chemotaxis and its influence on dispersal~\cite{taktikos2013motility} and clustering~\cite{engelmann1881neue}. For bacteria, this can be realized by genetically suppressing tumbling~\cite{NO_THUMBLE}. Our particles or cells divide at a nutrient-dependent rate, while divisions are assumed to be quasi-instantaneous. We consider a constant nutrient influx from a point source. The time scales of motility and proliferation are strongly separated for microorganisms such as \emph{E. coli}, which also requires large system sizes to explore the non-uniform population and nutrient distributions. Therefore, to keep the computational effort manageable, we squeeze the time scales together. A possible experimental realization is a highly viscous fluid which reduces cell motility~\cite{VISCOSITY_MOBILITY}. 

Within such a minimalistic model, the steady population of the bacteria only depends on the ratio between nutrient supply and bacterial death, as expected from a uniform model. In contrast, the nutrient amount is highly governed by its non-uniform distribution around the source, which is tunable by the nutrient diffusion coefficient. In turn, for passive and weakly active particles the nutrient distribution regulates the spatial extent of the bacterial cluster and thereby provides a clustering mechanism that does not rely on chemotaxis or adhesion. However, for high activity these clusters dissolve and the bacteria cover the entire system area even if the nutrient is localized. The transient growth towards a steady state is also heavily influenced by the nutrient diffusion coefficient. We show that it can be
mapped onto the dynamics of a uniform model using two limiting cases where we choose effective system sizes for the nutrient and bacterial distribution.

The article is structured as follows. In Sec.~\ref{sec:model} we introduce the particle-based and the uniform model. For immobilized bacteria or passive particles, we investigate the steady state in Sec.~\ref{sec:steady} and the transient growth dynamics in Sec.~\ref{sec:transient}. We compare our findings to the results for motile bacteria or active particles in Sec.~\ref{sec:active}. In the conclusions of Sec.~\ref{sec:conclusion} we summarize our results and present an outlook.

\section{Model}\label{sec:model}

In this section, we formulate a particle-based model for dividing bacteria consuming a nutrient, choose feasible bio-inspired parameters, and introduce a model that describes a uniform system as a reference.

\begin{figure*}
    \centering
    \includegraphics[scale=1]{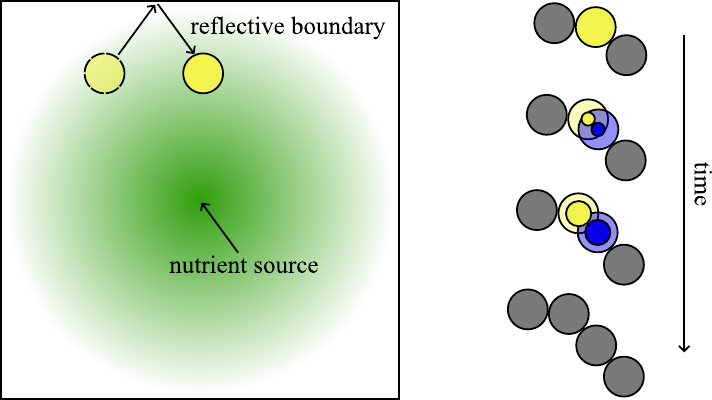}  
    \caption{(left) Sketch of the setup with a source of diffusing nutrients in the centre, reflective boundary conditions, and a quadratic box shape. (right) Mechanical implementation of the cell division. During the division, the circular parent cell is marked in yellow and the daughter cell is marked in blue. Their radii grow in time until the value $\sigma/2$ is reached.
    }\label{fig:model}
\end{figure*}

\subsection{Particle-based model}\label{sec:particle_based}
The setup of the two-dimensional system is sketched in Fig.~\ref{fig:model} (left). The bacteria are described as overdamped active Brownian particles experiencing thermal noise on their positions and orientations:
\begin{eqnarray}\label{eq:ABP}
    \dot{\mathbf{r}}_i &= v \mathbf{u}_i + \mu \sum_{i\neq j} \mathbf{F}_{ij} +  \sqrt{2D_T} \bm{\xi}_i \label{eq:dyn_position}\,, \nonumber \\
    \dot{\varphi}_i &= \sqrt{2D_R} \eta_i \label{eq:dyn_angle} \,,
\end{eqnarray}
with orientation vector $\mathbf{u}_i = (\cos\varphi_i, \sin\varphi_i)$, propulsion velocity $v$, and translational and rotational diffusion coefficients $D_T$ and $D_R$, respectively. For pure thermal noise they are linked to the respective translational and rotational mobilities $\mu_T$ and $\mu_R$ by $D_{T/R} = \mu_{T/R} k_B T$. For a spherical particle, the Stokes friction coefficients yield $D/D_R=\mu/\mu_R = \sigma^2/3$. The Gaussian noises $\xi_{\alpha,i},\eta_i$ are delta-correlated with zero mean and a variance of one. The volume exclusion force $\mathbf{F}_{ij} = \mathbf{F}_{\sigma}(\mathbf{r}_i-\mathbf{r}_j)$ resembles a purely repulsive spring with stiffness $k$:
\begin{eqnarray}\label{eq:force}
    \mathbf{F}_{\sigma}(\mathbf{r})= k \theta(|\mathbf{r}|-\sigma) (|\mathbf{r}|-\sigma) \hat{\mathbf{r}} \,.
\end{eqnarray}
The Heaviside step function $\theta(r)$ cuts off the interaction at the particle diameter $\sigma$. Particles are contained in the system by a reflective boundary condition to avoid accumulation at the wall and an additional repulsive spring interaction between particles and wall to prevent particles from being pushed out of the system area by other particles.

We introduce a two-dimensional nutrient concentration field $s(\mathbf{r},t)$. A nutrient is pumped into the system at rate $R_0$ at location $\mathbf{r}_s$, diffuses with diffusion coefficient $D_N$, and is eaten up by the bacteria:
\begin{eqnarray}\label{eq:nutrient_individual}
        \frac{\partial s(\mathbf{r},t)}{\partial t} = &D_N \nabla^2 s(\mathbf{r},t) + R_0\cdot\delta(\mathbf{r} - \mathbf{r}_s) -
        \sum\limits_{i=1}^{N}\delta(\mathbf{r}-\mathbf{r}_i(t))\gamma g(s(\mathbf{r}_i,t))\,.
\end{eqnarray}

The individual uptake rate $\gamma g(s(\mathbf{r}_i,t))$ is proportional to the nutrient-dependent growth rate $g(s)$ and the parameter $\gamma$ is the amount of nutrient required to make a new cell~\cite{BACTERIAL_GROWTH}. Nutrient amounts will be given in reference to this value. The growth rate follows the Monod function~\cite{monod1949growth}
\begin{eqnarray}\label{eq:monod}
    g(s)=\frac{g_\mathrm{max}s}{K_s+s} \,,
\end{eqnarray}
with maximal growth rate $g_\mathrm{max}$ and characteristic area concentration $K_s$, where the growth rate becomes $g_\mathrm{max}/2$. The area concentration is related to the volume concentration, $\mathcal{K}_s = K_s\sigma^{-1}$, used in real three-dimensional systems using the ``height'' $\sigma$ of our quasi-two-dimensional system.

To include the nutrient-dependent growth and starvation process, we introduce the rule:
\begin{center}\parbox{\dimexpr\linewidth-10\fboxsep-10\fboxrule\relax}{\textit{In a time interval $dt$, each particle has the probabilities of $dt \cdot g(s(\mathbf{r}_i,t))$ to divide and $dt \cdot d$ to die.}}\end{center}\label{box:test} 
Here, we choose a constant death rate $d$. If a parent cell divides, a daughter cell gets placed on the parent cell. They interact via a repulsive spring force $\mathbf{F}_{\sigma_d(t)}(\mathbf{r}_i-\mathbf{r}_j)$, introduced in Eq.~(\ref{eq:force}), with growing rest distance $\sigma_d(t) = v_d t$. The speed of the division process is regulated by $v_d$. If the growing rest distance reaches the full particle diameter $\sigma_d(t) = \sigma$ or the particles drift apart $|\mathbf{r}_i -\mathbf{r}_j| \geq \sigma$, the division process is finished. Parent and offspring now interact like all other particles via the force in Eq.~(\ref{eq:force}). The division process is sketched in Fig.~\ref{fig:model}\ (right). This implementation is similar to Ref.~\cite{li2021role}, however, we assume a constant cell size after the cell division is finished.

This model is an example of a hybrid particle-continuum reaction-diffusion system. For the simulation we rescale the equations as shown in~\ref{sec:rescales}.

\subsection{Time scales and simulation parameters}\label{sec:timescales_parameters}
\begin{figure*}
  \centering
  \begin{minipage}[c]{0.49\textwidth}
    \centering
    \includegraphics[scale=1]{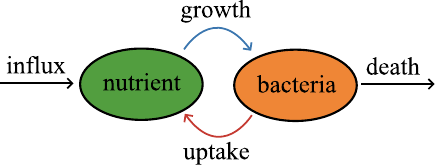}
  \end{minipage}
  \hfill
  \begin{minipage}[c]{0.49\textwidth}
    \centering
    \includegraphics[scale=1]{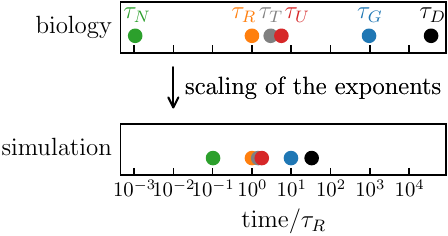}
  \end{minipage}
  \caption{(left) Reductionist sketch of the system. (right) Time scales of the biological system before scaling and time 
  scales of the simulation after scaling. The processes from left (fast) to right (slow) with their respective characteristic 
  time scales are: diffusion ($\tau_N$), bacterial motion ($\tau_R$, $\tau_T$), nutrient uptake ($\tau_U$), cell 
  growth ($\tau_G$), and cell death ($\tau_D$).}\label{fig:reduction}
\end{figure*}

A reductionist perspective on our system with relevant time scales is sketched in Fig.~\ref{fig:reduction}~(left): I) bacteria and II) nutrient couple via III) growth and IV) nutrient uptake. Additionally, there is V) a constant influx of nutrients and VI) an outflux of bacteria due to the death of individual cells. We define time scales for all six aspects:
\begin{itemize}
    \item[I)] bacterial motion: $\tau_R = D_R^{-1}$ - orientational correlation time and $\tau_T = \sigma^2 D_T^{-1}$ 
    - translational diffusion time
    with $D_T = \frac{\sigma^2}{3} D_R$ for spherical particles 
    using Stokes friction.
    \item[II)] nutrient diffusion: $\tau_N = \sigma^2 D_N^{-1} $
    - time it takes, for example, glucose to diffuse over the characteristic distance.
    \item[III)] bacterial growth: $\tau_G = g^{-1}(K_s) = 2g^{-1}_\mathrm{max}$ - inverse growth rate at concentration $K_s$.
    \item[IV)] nutrient uptake $\tau_U =2\sigma^2 K_s  \gamma^{-1} g_\mathrm{max}^{-1}$ - time to consume the nutrient of concentration $K_s$ in the local surrounding of the bacteria with area
    $\sigma^2$.
    \item[V)] nutrient influx: $\tau_I=\gamma R_0^{-1}$ - time to insert the nutrient amount necessary to make a new cell. 
    This can be easily varied in an experimental setup.
    \item[VI)] bacterial death: $\tau_D = d^{-1}$ - inverse death rate.
\end{itemize}
The biological system - \textit{Escherichia coli} in $\SI{41}{^\circ C}$ water with glucose as the limiting nutrient - dictates the parameters (\ref{sec:parameters}). The resulting time scales span over seven orders of magnitude, as shown in Fig.~\ref{fig:reduction} (right). In our particle-based simulation, this is not feasible. Therefore, we squeeze the time scales together such that the ratios of their orders of magnitude (in units of $\tau_R$), defined as the logarithms, are preserved (e.g. $\log{\frac{\tau_G}{\tau_R}}/\log{\frac{\tau_U}{\tau_R}} = \text{const}$). More concretely, we reduce all the orders of magnitudes by a factor of $3$. The procedure is demonstrated for the time scale of bacterial growth:
\begin{eqnarray}
    \tau_G = \SI{952.4}{\tau_R} =10^{2.979}\,\si{\tau_R} \to \tau_G = 10^{2.979/3}\,\si{\tau_R} = \SI{9.839}{\tau_R} \,.
\end{eqnarray}
The transformed time scales in Fig.~\ref{fig:reduction} (right) span less than three orders of magnitude and can be realized in the simulation. 

We recover the simulation parameters from the transformed time scales $g_\mathrm{max} = 2\tau_G^{-1} = \SI{0.203}{\tau_R^{-1}}$, $d= \tau_D^{-1} = \SI{0.030}{\tau_R^{-1}}$, $K_s = \gamma\sigma^{-2} \tau_U\tau_G^{-1} = \SI{0.181}{\gamma\sigma^{-2}}$, $D_N = \sigma^{-2} \tau_N^{-1}= \SI{9.775}{\sigma^2\tau_R^{-1}}$, and $D_T = \sigma^2\tau_T^{-1}=\SI{0.693}{\sigma^2 \tau_R}^{-1}$. We additionally vary the propulsion velocity $v$ by choosing a persistence number $\mathrm{P} =  v \sigma^{-1}\tau_R$ between $\mathrm{P}=0$ and $20$, that corresponds to velocities up to $v = \SI{10.562}{\micro m s^{-1}}$. We also implement a sufficiently large spring constant $k = \SI{3e4}{k_B T \sigma^{-2}}$. Furthermore, we set the division speed $v_d = 20\sigma\tau_R^{-1}$, which guarantees that the time scale of division $\sigma v_d^{-1}$ is much faster than the time scale of bacterial growth $\tau_G$. The simulation box is quadratic with edge length $L_x=L_y=\SI{200}{\sigma}$. The nutrient source is placed in the centre. The setup is sketched in Fig.~\ref{fig:model} (left). At the start of the simulation, $N_0 = 0.1\cdot N^*$ bacteria are placed on a square grid and the nutrient field is set to $s=0$. Here, $N^*$ is the steady-state population in the uniform model introduced in Sec.~\ref{sec:uniform}. The simulation time is $\SI{1500}{\tau_R}$ and properties concerning the stationary state use the data between $\SI{1000}{\tau_R}$ and $\SI{1500}{\tau_R}$.

\subsection{Uniform model}\label{sec:uniform}
As a reference, we introduce a scalar model assuming that nutrients and bacteria are uniformly distributed. When comparing to the simulations, it is of advantage to introduce different areas: the area $A_S$ occupied by the nutrients and the area $A_N \geq A_S$ of the bacteria. Then, the dynamics of population number $N$ and the total nutrient amount $S$ is described as
\begin{eqnarray}\label{eq:no_space}
        \dot N &= \frac{A_S}{A_N}g(S/A_S)N - Nd \,, \nonumber \\ 
        \dot S &=R_0 - \gamma \frac{A_S}{A_N} g(S/A_S) N \,.
\end{eqnarray}
We assume here a constant influx of nutrient, while outflux of bacteria is only caused by bacterial death. The ratio $s=S/A_S$ indicates nutrient density. The ratio $A_S/A_N \leq 1$ comes in since only a fraction of the bacteria, $N A_S /A_N$, has access to nutrient and thereby consume them and grow. The case $A_N = A_S$ presents a limiting case of the population model in Ref.~\cite{khatri2012oscillating} and is distinct from the chemostat, where bacterial and nutrient depletion is caused by material outflux~\cite{CHEMOSTATEN_THEORIE}. Because the ratio $A_S/A_N$ can be interpreted as decreasing the growth rate $g$, we introduce the effective growth rate $\mathcal{G}(s) = \frac{A_S}{A_N}g(s) = \frac{\mathcal{G}_\mathrm{max} s}{s+K_s}$  with the reduced maximum growth rate $\mathcal{G}_\mathrm{max} =  \frac{A_S}{A_N} g_\mathrm{max}$. The fixed point of the dynamic equations,
\begin{eqnarray}\label{eq:uniform_fixedpoint}
    N^* = \frac{R_0}{d \gamma} \hspace{5mm}  \text{and} \hspace{5mm} S^* = A_S \mathcal{G}^{-1}(d) \,,
\end{eqnarray}
gives the steady population $N^*$ and nutrient amount $S^*$. Note that $N^*$ does not depend on system size. This is plausible because, in steady state the bacterial outflux rate due to death, $d N^*$, and bacterial influx rate due to nutrient-induced division, $R_0/ \gamma$, need to balance each other for all system sizes. 

\begin{figure*}
    \hspace*{\fill}  
    \includegraphics[scale = 1.2 ]{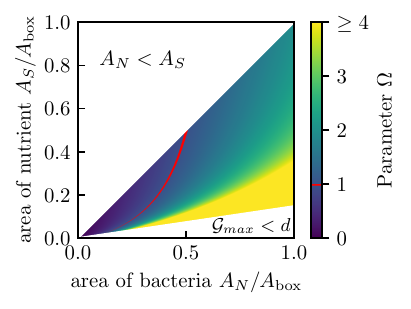}
    \hspace*{\fill}  
    \caption{Parameter $\Omega$ over nutrient area $A_S$ and bactrial area $A_N\geq A_S$. For $\mathcal{G}_\mathrm{max}<d$ death dominates growth and the population dies out. Parameters are chosen as discussed in Sec.~\ref{sec:timescales_parameters} with steady population $N^*=3000$.}\label{fig:omega}
\end{figure*}

To characterize the fixed point, we evaluate the eigenvalues of the Jacobian,
\begin{eqnarray}\label{eq:eigenvalues}
    \lambda_{1,2} = \frac{2d}{\Omega} \left[-1 \pm \sqrt{1-\Omega}\right] \, ,
\end{eqnarray}
where the parameter
\begin{eqnarray}\label{eq:omega}
    \Omega = \frac{A_S}{A_c N^*} \hspace{5mm} 
    \text{with} \hspace{5mm} A_c= \frac{\gamma ( \mathcal{G}_\mathrm{max} -d )^2}{4 d\mathcal{G}_\mathrm{max} K_s}
\end{eqnarray}
relates the nutrient area $A_S$ to the critical area $A_c N^*$. The fixed point is always stable because $\mathrm{Re}(\lambda_{1,2}) < 0$, but the critical nutrient area $A_S = A_c N^*$ corresponding to $\Omega = 1$ separates overdamped ($\Omega<1$) and damped oscillatory ($\Omega>1$) population dynamics from each other. In Fig.\ \ref{fig:omega} we illustrate the dependence of $\Omega$ on $A_S$ and $A_N$. The red line indicates $\Omega=1$. While for $A_S = A_N$ the dynamics switches between overdamped and damped oscillatory, the parameter $\Omega$ is always larger than 1 to the right of the red line, for example, for $A_N = A_\mathrm{box}$.

\begin{figure*}
    \hspace*{\fill}  
    \includegraphics[scale=1]{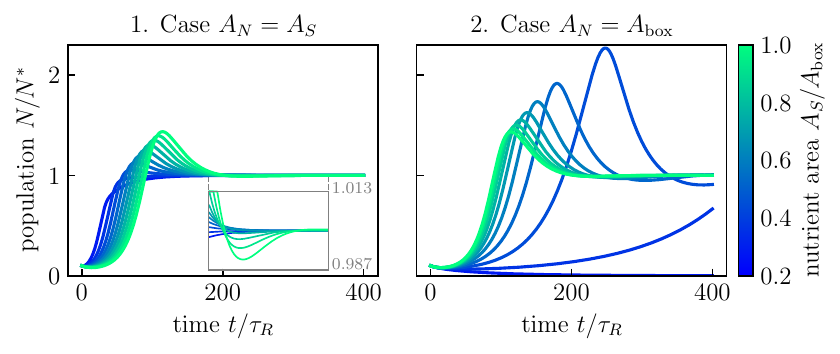}
    \hspace*{\fill}  
    \caption{Population dynamics for (left) case 1 with $A_N = A_S$ and (right) case 2 with $A_S< A_N = A_\mathrm{box}$ for varying $A_S$. Areas are given in units of the system area $A_\mathrm{box}$. In both cases a characteristic overshoot of the bacterial population is observed. Parameters are chosen as discussed in Sec.~\ref{sec:timescales_parameters} with steady population $N^*=3000$.
    }\label{fig:uniform_model}
\end{figure*}

In the following, two cases will be relevant: (1) nutrient and bacteria cover the same area $A_N = A_S$ and (2) bacteria cover the entire system area $A_N=A_\mathrm{box}$ while nutrient can be localized with $A_S\leq A_\mathrm{box}$. We will use case 1 to fit the simulation data of passive particles, while case 2 applies to highly active particles that always spread over the whole system due to their activity.Figure\ \ref{fig:uniform_model} illustrates that the population dynamics of the two cases show a different trend when the area of nutrient $A_S$ is increased. For $A_N=A_S$ (left) an increase of $A_S$ increases the characteristic overshoot in $N$, because the lower densities result in a weaker coupling between nutrient and bacterial population. For $A_S \le A_N = A_\mathrm{box}$ (right), the characteristic overshoot in $N$ always exists, but now decreases with an increase of $A_S$ because a larger fraction of the bacteria can interact with the nutrient and coupling is stronger. Furthermore, with decreasing $A_S$ the dynamics towards steady state is considerably slowed down.

\section{Results for Passive Particles}
\begin{figure*}
    \hspace*{\fill}  
    \includegraphics[scale = 1]{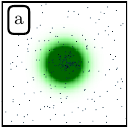}
    \includegraphics[scale = 1]{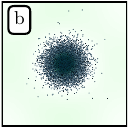}
    \includegraphics[scale = 1]{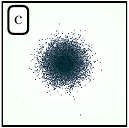}
    \hspace*{2.4mm}  
    \includegraphics[scale = 1]{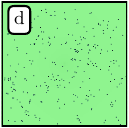}
    \includegraphics[scale = 1]{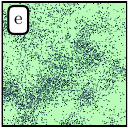}
    \includegraphics[scale = 1]{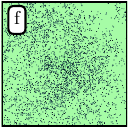}
    \hspace*{\fill} 
    \\
    \hspace*{\fill}  
    \includegraphics[scale = 1]{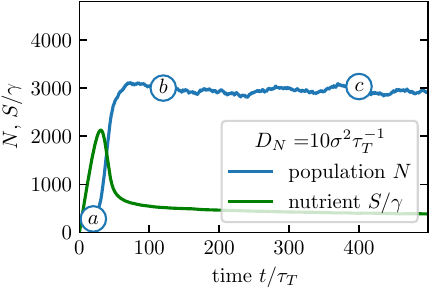}
    \hspace*{2mm}  
    \includegraphics[scale = 1]{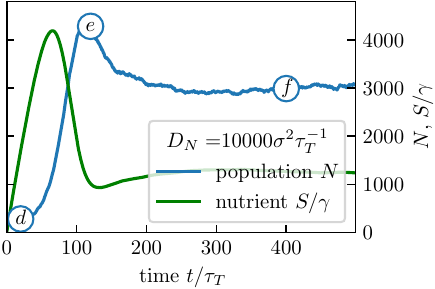}
    \hspace*{\fill}  
    \caption{(Bottom) Dynamics of population $N$ and nutrient amount $S$ with nutrient input rate $R_0 = 3000 d\gamma$ for the diffusion coefficients $D_N = \SI{10}{\sigma^2\tau_R^{-1}}$ (left) and $D_N = \SI{10000}{\sigma^2\tau_R^{-1}}$ (right). (Top) The snapshots show different stages of the growth process as indicated in the bottom plots.  Dots represent particles and the green shading the nutrient field. For small $D_N$ the nutrient is almost entirely covered by particles in the steady state.}\label{fig:dynamics}
\end{figure*}

We start our investigations with passive particles ($\mathrm{P}=0$), which helps us to work out the effect of activity or self-propulsion in Sec.\ \ref{sec:active}. Figure~\ref{fig:dynamics} gives an overview of our essential results from the 2D simulations. For a small nutrient diffusion coefficient $D_N=\SI{10}{\sigma^2\tau_R^{-1}}$ (left), the population $N$ grows continuously until the steady state is reached, while the nutrient amount $S$ goes through a maximum. The population strongly clusters around the source of the nutrient, which is nonuniformly distributed. In contrast, for large $D_N =\SI{10000}{\sigma^2\tau_R^{-1}}$ (right), the nutrient spreads evenly over the whole system area and the population is almost uniform during the growth process. The population relaxes faster towards the steady state for the small $D_N$, but the steady population is not affected by the nonuniform distribution of nutrient and particles.

In the following, we will establish a connection between the population dynamics and heterogeneities in the system. We will introduce an effective system size within the uniform model of Eq.~(\ref{eq:no_space}) to understand the role of such heterogeneities. Following the observation that nutrient and bacteria cover roughly the same area, we equate both areas in the uniform model, $A_N=A_S$. This will give us insights into the steady state (Sec.~\ref{sec:steady}) and the transient phase (Sec.~\ref{sec:transient}). The parameters not explicitly mentioned are chosen as discussed in Sec.~\ref{sec:timescales_parameters}.

\subsection{Steady State}\label{sec:steady}
\begin{figure*}
    \hspace*{\fill} 
    \includegraphics[scale=1]{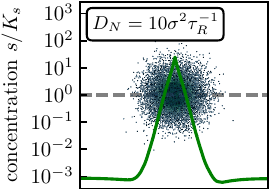}
    \includegraphics[scale=1]{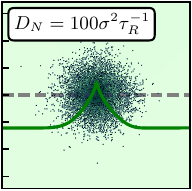}
    \includegraphics[scale=1]{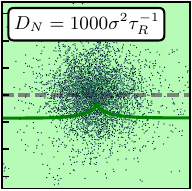}
    \includegraphics[scale=1]{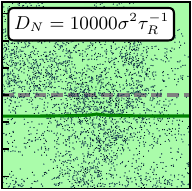}
    \hspace*{\fill} 
    \caption{Particle distribution in the steady state for different $D_N$ and with $N^*=3000$. The green shading shows the nutrient distribution. The green curve illustrates the nutrient concentration along the grey dashed line. The nutrient profile gets flatter for larger $D_N$.}\label{fig:snapshot}
\end{figure*}

We start with analysing the steady state. Figure~\ref{fig:snapshot} shows snapshots of the particle distribution with different degree of heterogeneity tuned by the nutrient diffusion coefficient $D_N$. For small $D_N$ the nutrient distribution is strongly peaked and overlaid by the strongly clustered particles that only occupy a fraction of the system area. Increasing $D_N$ allows the population to spread over larger areas. For $D_N=\SI{10000}{\sigma^2\tau_R^{-1}}$ the cluster has completely dissolved and the entire area is covered by particles. In other words there is an effective system area $A_\mathrm{eff}$ occupied by the particles and nutrient that grows with the nutrient diffusion coefficient and reaches the total system area for sufficiently large $D_N$.

\begin{figure*}
    \begin{center}
        \includegraphics[scale = 0.98]{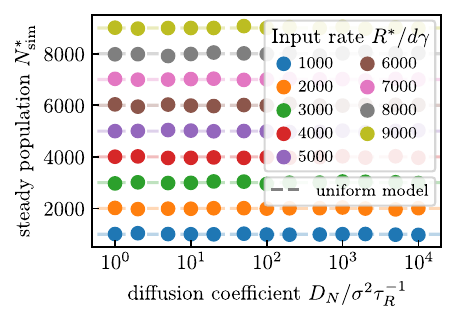}
        \includegraphics[scale = 0.98]{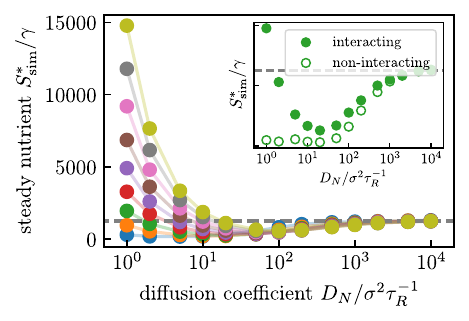} 
    \end{center}
    \caption{
    (left) Steady population of the particle-based model, $N^*_\mathrm{sim}$, and (right) steady nutrient amount $S^*_\mathrm{sim}$ (in units of characteristic nutrient amount $\gamma$) plotted versus nutrient diffusion coefficient $D_N$ for different nutrient input rates $R_0$ (in units of $d \gamma$). Dashed lines indicate the predictions of the uniform model, $N^*$ and $S^*$, as stated in Eqs.~(\ref{eq:uniform_fixedpoint}) and the full lines in the right plot are guides to the eye. The inset compares steady nutrient amount versus $D_N$ for particles with and without hard-core repulsion for $R_0 = 3000 d \gamma$.
    }\label{fig:steady_state}
\end{figure*}

First, we look at population $N^*_\mathrm{sim}$ and nutrient amount $S^*_\mathrm{sim}$ in the steady state, which we determined in the 2D simulations. Figure~\ref{fig:steady_state} shows both quantities plotted versus $D_N$ for different nutrient input rates $R_0$. The steady state of the uniform model in Eq.~(\ref{eq:uniform_fixedpoint}) with $N^* = R_0 /(d \gamma)$, $S^* = A_S \mathcal{G}^{-1}(d)$, and $A_S = A_\mathrm{box}$ is represented by the dashed lines. Most notably, the population in the steady state, $N^*_\mathrm{sim}$, is independent of $D_N$ and, thus, it is not influenced by the degree of heterogeneity, for example, in the nutrient distribution. The argument used in Sec.~\ref{sec:uniform} explaining the independence of $N^*$ from system size also applies here. Bacterial influx $R_0/\gamma$ and bacterial outflux $dN^*$ always need to balance for all heterogeneities and system sizes. In contrast, the nutrient amount strongly depends on $D_N$ and only for $D_N/\si{\sigma^2 \tau_R^{-1}} \to \infty$ converges to the prediction of the uniform model, for which we used the system size, $A_S=A_\mathrm{box}$. This is the case, because for large $D_N$ the nutrient is distributed almost uniformly and the uniform model applies with $A_S=A_N=A_\mathrm{box}$.

Since $S^\ast \propto A_S$, one could be tempted to introduce an effective system size $A_\mathrm{eff}$ to explain the results from the particle-based simulation in Fig.\ \ref{fig:steady_state}, right, which deviate from the prediction of the uniform model for small and intermediate $D_N$. However, we already know that the occupied area $A_\mathrm{eff}$ grows with $D_N$ and, therefore, would expect $S^*_\mathrm{sim}$ to grow with $D_N$. But this is only the case for $D_N$ larger than ca. $\SI{e2}{\sigma^2\tau_R^{-1}}$, for smaller $D_N$ the nutrient amount in the steady state rises again.

We propose that this is caused by the limited uptake rate per area the particles can provide. We consider the specific case $R_0 = 3000 d \gamma = \SI{90}{\tau_R^{-1}\gamma}$. The diffusion equation Eq.~(\ref{eq:nutrient_individual}) of the nutrient is solved on a lattice with constant $2\sigma$. Thus, the nutrient from the delta-peaked source is initially distributed over the area $4\sigma^2$, which results in an input rate per area of $\SI{22.5}{\tau_R^{-1}\sigma^{-2}\gamma}$. In contrast, the maximum uptake rate per individual, $\gamma g_\mathrm{max}$, and the maximum number density in hexagonal tight packing, $\rho_\mathrm{max}=\frac{2}{\sqrt{3}} \sigma^{-2} \approx \SI{1.15}{\sigma^{-2}}$~\cite{chang2010simple}, limits the maximum uptake rate per area to $\gamma g_\mathrm{max} \rho_\mathrm{max} \approx 0.234 \gamma \tau_R^{-1}\sigma^{-2}$. This value is much lower than the input rate per area at the position of the source. The nutrient needs to diffuse away before it can entirely be consumed, which needs more time for smaller $D_N$. As a consequence, the nutrient amount in the steady state rises. To validate this hypothesis, we performed simulations with particles, which do not interact by volume exclusion. Thus, the maximum density and thereby the maximum uptake rate per area is no longer limited. Indeed, the inset in Fig.~\ref{fig:steady_state}, right shows that now the steady nutrient amount $S^*_\mathrm{sim}$ is monotonously rising with $D_N$, as predicted.

\subsection{Transient dynamics}\label{sec:transient}
\begin{figure*}
    \centering
    \begin{subfigure}[b]{0.52\textwidth}
    \includegraphics[scale=1]{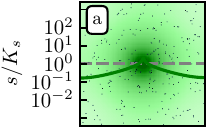}
    \includegraphics[scale=1]{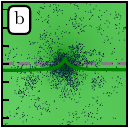}
    \includegraphics[scale=1]{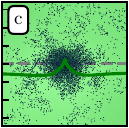}\\
    \includegraphics[scale=1]{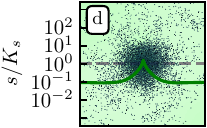}
    \includegraphics[scale=1]{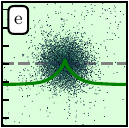}
    \includegraphics[scale=1]{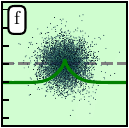}
    \end{subfigure}
    \begin{subfigure}[b]{0.46\textwidth}
    \includegraphics[scale=1]{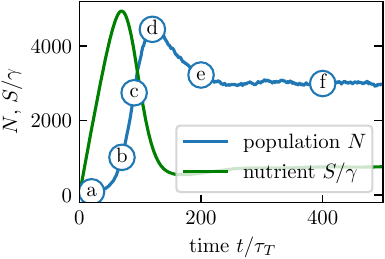}
    \end{subfigure}
    \caption{Transient dynamics of a system with steady population $N^* =3000$, initial population $N_0 = 100$, and diffusion 
    coefficient $D_N = \SI{200}{\sigma^2\tau_R}$.
    Left: Snapshots of evolving particle population and nutrient distribution (green shading) at different times as indicated in the right plot. The green curve illustrates the nutrient concentration along the grey dashed line. Right: Population $N$ and nutrient amount  $S$ plotted versus time.
    }\label{fig:snapshot_transient}
\end{figure*}

Now that we established a first connection between nutrient diffusion coefficient $D_N$, the nonuniform steady-state nutrient distribution, and the area occupied by the particles, we proceed by looking at the transient dynamics. First, we look closer at the spatial distribution of the growing and evolving particle population. The snapshots in Fig.~\ref{fig:snapshot} convey that in steady state the particle distribution is radially symmetric. This is certainly not the case in the beginning of the growth process as the snapshot (a) in Fig.~\ref{fig:dynamics} already shows. In Fig.~\ref{fig:snapshot_transient} we consider the system dynamics  for $N^* = 3000$ and $D_N = \SI{200}{\sigma^2\tau_R}$. However, to amplify the effect of statistical fluctuations, a smaller initial population $N_0 = 100$ is chosen in contrast to $N_0 = N^* /10$ used otherwise. The simulation starts with a small population (a). The nutrient quickly builds up until it reaches a maximum (b); now particle growth is possible in large parts of the system. Individual particles establish clusters across the entire system, the most prominent around the nutrient source (c). Because stochastic fluctuations strongly affect the distribution of the few initial particles, the clusters are unevenly distributed and also the nutrient distribution becomes slightly asymmetric. The clusters grow, consume nutrients, and overshoot the carrying capacity of the steady state (d). In parallel, a depletion zone of nutrients around the central cluster develops (c). As a consequence of the nutrient depletion, the clusters in the outer regions start to dissolve (e). As the system relaxes further toward the steady state, the outer clusters completely dissolve, the central cluster becomes symmetric, and the depletion zone vanishes (e,f).

For larger $D_N$ the asymmetry tends to be more prominent, while for small $D_N$ only the centre is covered by nutrients and clusters cannot grow in the outer regions. However, the nutrient depletion zone is more pronounced for small $D_N$, as Fig.~\ref{fig:dynamics} clearly shows. The diffusion coefficient $D_N = \SI{200}{\sigma^2\tau_R}$ depicted in Fig.~\ref{fig:snapshot_transient} is an intermediate value, where both effects can be observed.

\begin{figure*}
    \centering
    \includegraphics[scale=0.95]{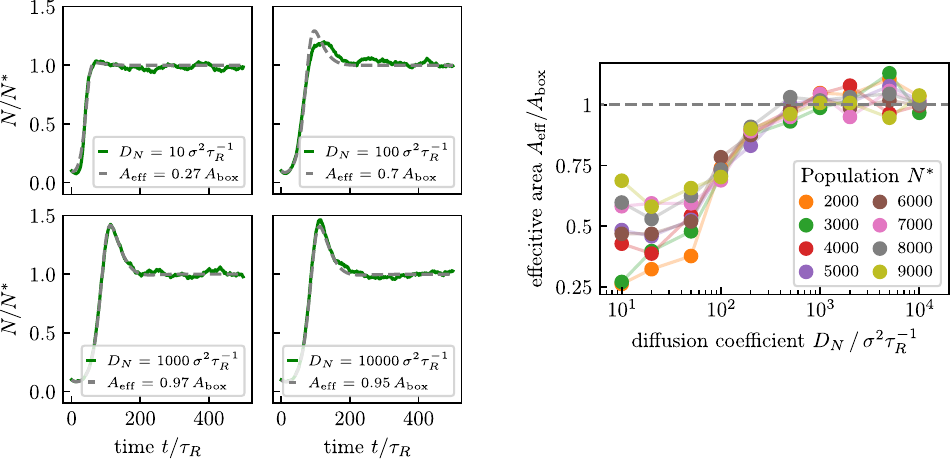}
    \caption{(left) Population dynamics of the particle-based simulations (green) with $N^*=3000$ for different diffusion coefficients of the nutrient, $D_N$. The dynamics is fitted with the uniform model of Eq.~(\ref{eq:no_space}) assuming $A_N=A_S=A_\mathrm{eff}$ (grey) using $A_\mathrm{eff}$ as a fit parameter. (right) Relation between the diffusion coefficient of nutrient $D_N$ and the effective system area $A_\mathrm{eff}$ for different steady populations $N^*$.}\label{fig:fits}
\end{figure*}

We further rationalize the relation between population dynamics and nonuniform nutrient distribution. In Fig.~\ref{fig:fits} (left) we plot population $N$ versus time for different $D_N$. The population dynamics is more strongly damped for more heterogeneous nutrient distributions, \emph{i.e.}, when $D_N$ is small. However,  for increasing $D_N$ we see that a clear overshoot in  $N$ develops. We did already observe the same transition from overdamped relaxation to damped oscillations in the uniform model for increasing and equal areas covered by nutrient and bacteria, $A_S = A_N$, as shown in Fig.~\ref{fig:uniform_model} (left). In the following we will introduce an effective system size $A_\mathrm{eff} = A_S=A_N$ to model this transition in the particle-based simulations and thereby account for heterogeneities in the system.
 
The grey dashed lines in Fig.~\ref{fig:fits} (left) show fits of the simulation results using the uniform model of Eq.~(\ref{eq:no_space}). To fit the curves, we use the fit parameter $A_\mathrm{eff} = A_N=A_S$ that can deviate from the simulation box size $A_\mathrm{box}$. For $D_N/\si{\sigma^2\tau_R^{-1}} = 10,1000,10000$ the fit captures the dynamics well, only for $D_N = \SI{100}{\sigma^2\tau_R^{-1}}$ a clear deviation is apparent. So, by choosing an effective system size $A_\mathrm{eff}$ that accounts for the fact that particles and nutrient display a nonuniform distribution in the simulation box, we are able to model the simulated dynamics. 

The fitted effective system area $A_\mathrm{eff}$ over the diffusion coefficient is shown in Fig.~\ref{fig:fits}~(right) for different steady populations $N^*$. We observe a monotonous relation between heterogeneity ($D_N$) and effective system size. In particular, for smaller $N^\ast$ the effective system size jumps at around $D_N=100 \sigma^2 \tau_R^{-1}$, while for large $D_N$ it approaches the total system area. This all agrees well with our former observation that for small $D_N$ the particle cluster only occupies a fraction of the total system, while for large $D_N$ it covers the entire area.

\section{Results for Active Particles}\label{sec:active}

So far we looked at immobilized bacteria modelled as passive Brownian particles. In this part we demonstrate how motility changes both steady-state distribution and population dynamics. We model motile bacteria as active Brownian particles using Eq.~(\ref{eq:ABP}). First, the case of high activity with propulsion velocity $v = \SI{20}{\sigma\tau_R^{-1}}$ corresponding to a persistence number $\mathrm{P}=20$ is explored. We can still describe its population dynamics with the uniform model with modified choices for the areas $A_N$ and $A_S$.Second, we consider a lower activity with $v = \SI{10}{\sigma\tau_R^{-1}}$ and $\mathrm{P}=10$. Here, the population dynamics is intermediate between the highly active and the passive case.

\subsection{High Activity $\mathrm{P}=20$}
\label{sec:highly_active}

\begin{figure*}
    \centering
    \includegraphics[scale=0.95]{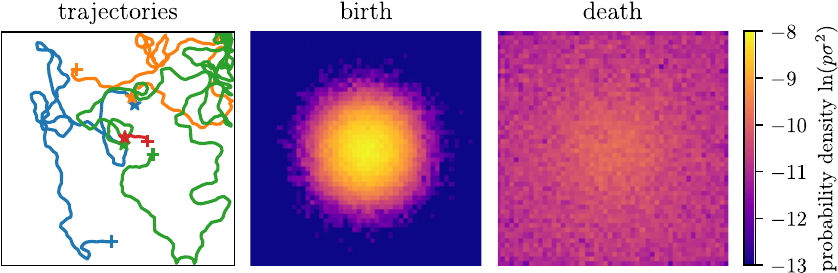}
    \caption{(right) Space-time tree of a lineage starting with one active particle. The stars and crosses indicate cell division and death, respectively. (centre) Spatial probability distribution of birth/cell divisions and (right) particle death. The data is obtained for active particles with persistence number $\mathrm{P}=20$, steady population $N^*=3000$, and 
    a small nutrient diffusion coefficient $D_N=\SI{10}{\sigma^2\tau_R^{-1}}$.}\label{fig:trajectories}
\end{figure*}

Figure~\ref{fig:trajectories} (left) shows the motion of a particle (blue) and all its descendants. The space-time trajectories known for conserved particle number are replaced by space-time trees of lineages of particles being born ($*$) and dying ($+$) as discussed in Ref.\ \cite{hallatschek2023proliferating}. The lineage depicted starts with one particle (blue trajectory) having two direct offspring (orange, green)  with one of them having another offspring (red). Then, the lineage goes extinct. We note when nutrient is localized around the centre, the birth of active particles is also limited to the central region, but their motility allows the particles to explore the area and die more evenly distributed over the system. This is illustrated in Fig.\ \ref{fig:trajectories} (centre, right).

\begin{figure*}
    \hspace*{\fill} 
    \includegraphics[scale=1]{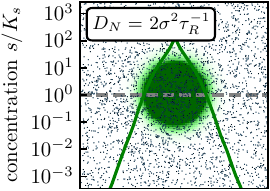}
    \includegraphics[scale=1]{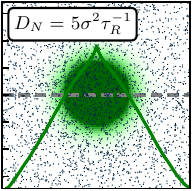}
    \includegraphics[scale=1]{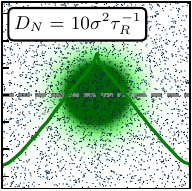}
    \hspace*{\fill} 
    \\
    \hspace*{\fill} 
    \includegraphics[scale=1]{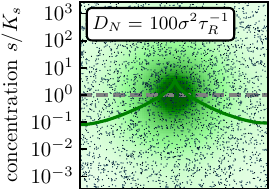}
    \includegraphics[scale=1]{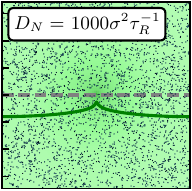}
    \includegraphics[scale=1]{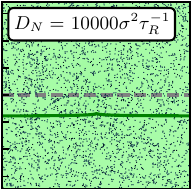}
    \hspace*{\fill} 
    \caption{Distribution of active particles and nutrient in the steady state $N^*=3000$. Particle velocity is  
    $v=\SI{20}{\sigma\tau_R^{-1}}$ ($\mathrm{P} = 20$)
    and $N^*=3000$. The green shading shows the nutrient distribution and the green curve illustrates the nutrient 
    concentration along the grey dashed line.
    }\label{fig:active_snap}
\end{figure*}

This feature is also apparent in Fig.~\ref{fig:active_snap}, where we show the snapshots of the steady state for different diffusion coefficients $D_N$ for $N^*=3000$. In strong contrast to the proliferation-induced clusters observed for passive particles in Fig.~\ref{fig:snapshot}, the distribution of highly active particles is almost uniform for all $D_N$, with slightly higher particle densities around the source for low $D_N$. For small $D_N$ the nutrient is localized, but the peak is less pronounced than for passive particles. This can be explained by the absence of the proliferation-induced particle cluster around the nutrient, which decreases the immediate uptake and allows the nutrient to diffuse further away from the source.

\begin{figure*}
    \hspace*{\fill} 
    \includegraphics[scale=1]{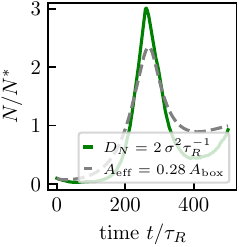}
    \includegraphics[scale=1]{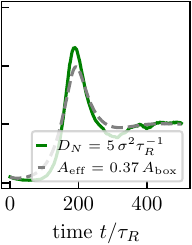}
    \includegraphics[scale=1]{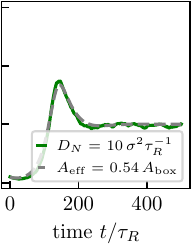}
    \vspace{3mm}
    \hspace*{\fill} 
    \\
    \hspace*{\fill} 
    \includegraphics[scale=1]{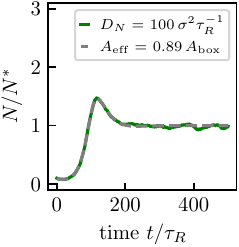}
    \includegraphics[scale=1]{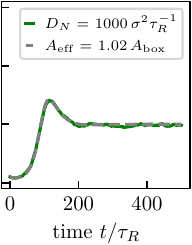}
    \includegraphics[scale=1]{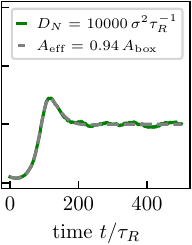}
    \hspace*{\fill} 
    \caption{Population curves of highly active particles for different $D_N$ with propulsion velocity $v=\SI{20}{\sigma\tau_R^{-1}}$ ($\mathrm{P} = 20$) and $N^*=3000$. The fitted curves (gray dashed lines) use Eq.~(\ref{eq:no_space}) by setting $A_N = A_\mathrm{box}$ and using $A_\mathrm{eff}=A_S$ as fit parameter.}\label{fig:active_fit}
\end{figure*}

The population dynamics of these highly active particles is presented in Fig.~\ref{fig:active_fit} and shows a different trend than for passive particles, which we addressed in Fig.~\ref{fig:fits} (left). While for passive particles the population overshoot increases with $D_N$, for active particles the overshoot is more pronounced for small $D_N$. The same two trends are observed in the scalar population model. We obtained them in Fig.~\ref{fig:uniform_model} by setting either $A_N = A_S$ or $A_S \le A_N = A_\mathrm{box}$.

How can we transfer this insight to the particle-based model? The assumption made for passive particles that bacteria and nutrient cover roughly the same areas, $A_N=A_S$, does not hold for active particles. Instead, we use the observation that bacteria spread approximately evenly over the entire system area and set $A_N = A_\mathrm{box}$, while we let the area of the nutrient $A_S$ vary with the diffusion coefficient $D_N$. Now, if nutrient is more localized, a smaller fraction of the bacteria interacts with the nutrient. Thus, the coupling between both species is weaker and the overshoot in the particle population is stronger.

To further validate this scenario for active particles, we fit the population model to the population curves obtained from the simulations using the area of the nutrient $A_S = A_\mathrm{eff}$ as the fit parameter. The resulting curves, gray dashed lines in Fig.\ \ref{fig:active_fit}, show good agreement with the particle-based simulation data. In particular, the time positions of the overshoots are always well captured. The fitted area of the nutrient $A_\mathrm{eff}$ grows with the diffusion coefficient and approaches $A_\mathrm{box}$ for larger $D_N$ as we expected. Only for low diffusion coefficients $D_N = 2$ and $5\,\si{\sigma^2\tau_R^{-1}}$ significant deviations are observed around the overshoots. Here, transient clusters occur and therefore the assumption that particles are evenly distributed does no longer apply. The forming and dissolving clusters also result in a time-dependent effective size $A_N$ of the bacterial distribution, which is not included in the model. Nonetheless, the tendency of a stronger overshoot for smaller $D_N$, the time positions of the overshoots, and the monotonous relationship between $D_N$ and $A_S$ are all well captured.

\subsection{Low Activity $\mathrm{P}=10$}\label{sec:weakly_active}

\begin{figure}
    \hspace*{\fill} 
    \includegraphics[scale=1]{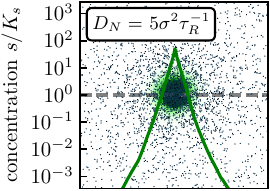}
    \includegraphics[scale=1]{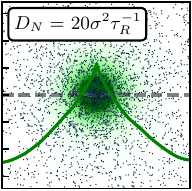}
    \includegraphics[scale=1]{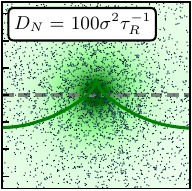}
    \includegraphics[scale=1]{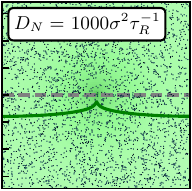}
    \hspace*{\fill} \\
    \hspace*{\fill} 
    \hspace*{1.2mm}
    \includegraphics[scale=1]{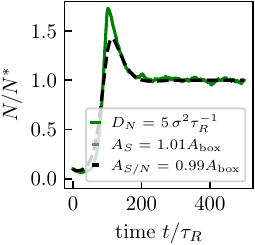}
    \includegraphics[scale=1]{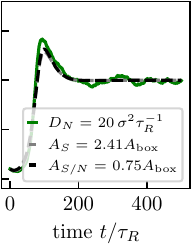}
    \includegraphics[scale=1]{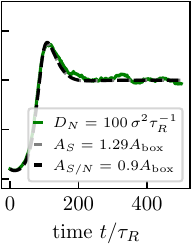}
    \includegraphics[scale=1]{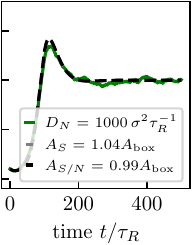}
    \hspace*{\fill} 
    \caption{Top: Snapshots for active particles with lower activity ($\mathrm{P}=10$) for different $D_N$ with $N^*=3000$. Bottom: Population dynamics. The gray dashed lines show fits with the uniform model with $A_N=A_{\mathrm{box}}$ as for highly active particles, while the black dashed lines use $A_N=A_S$ as in the passive case.
    }\label{fig:low_active}
\end{figure}

So far we discussed passive and strongly active particles with distinct collective behaviour and population dynamics. Now the intermediate case of low activity (persistence number $\mathrm{P}=10$) is illuminated. Snapshots of the steady state in Fig.~\ref{fig:low_active} (top) show that for small $D_N$ clusters form, but are less pronounced compared to the passive case in Fig.~\ref{fig:snapshot}, and a significant fraction of particles is outside the central cluster. Clusters dissolve for intermediate diffusion coefficients, e.g., $D_N=\SI{100}{\sigma^2 \tau_R^{-1}}$ but the bacterial density is still higher around the source.

The transient population dynamics is shown in Fig.~\ref{fig:low_active} (bottom). Similar to the highly active case, the population overshoot is more pronounced for small $D_N$. Fitting the observed dynamics with the uniform model and $A_N = A_\mathrm{box}$ as before (gray dashed line), works very well for the largest $D_N=\SI{1000}{\sigma^2 \tau_R^{-1}}$. However, already at $D_N=\SI{100}{\sigma^2 \tau_R^{-1}}$ and very pronouncedly at $D_N=\SI{20}{\sigma^2 \tau_R^{-1}}$ the fitted area for the nutrient, $A_S > A_\mathrm{box}$, is un\-physical. The reason is that bacteria are localized such that the premise of evenly distributed bacteria in the whole system, $A_N = A_\mathrm{box}$, which we used for highly active particles, does not apply. As an alternative, we treated $A_S$ and $A_N$ as independent parameters. However, this generated ambiguous parameter pairs. In the end we tried to fit with $A_S=A_N$ (black dashed lines), which gives realistic values $A_S=A_N < A_\mathrm{box}$ and a very good fit for $D_N=\SI{100}{\sigma^2 \tau_R^{-1}}$. The fit clearly does not reproduce the height of the overshoot for $D_N=\SI{5}{\sigma^2 \tau_R^{-1}}$ since the assumption of a uniform distribution of the active particles is strongly violated.

\section{Conclusion}\label{sec:conclusion}
We presented a minimalistic model for bacterial growth in a nonuniform nutrient environment, where we couple proliferating active or passive Brownian particles to a diffusing nutrient emitted from a point source. This allows us to gain insights into the interplay between collective behaviour and population dynamics of bacteria. The heterogeneity of the nutrient distribution around the point source highly depends on its diffusion coefficient $D_N$. To keep the computational effort manageable, we needed to squeeze the time scales of motility and growth together.

Using numerical simulations, we show that localizing the nutrient supply limits the effective area of bacterial growth and results in proliferation-induced bacterial clustering around the nutrient source for passive and weakly active particles. This presents an alternative clustering mechanism that does not rely on chemotaxis or adhesion. In contrast, highly active particles can disperse over the whole system during their lifetime resulting in a nearly uniform distribution. Interestingly, the space-time trajectories of non-dividing particles are replaced by space-time trees of lineages that can persist, flourish, and die out~\cite{hallatschek2023proliferating}. For both active and passive particles the steady population is purely determined by the ratio of nutrient influx to cell death and can be predicted by a uniform model.

The transient dynamics of the system strongly depends on the nutrient and bacterial distribution, which are determined by the nutrient diffusion coefficient and bacterial motility or activity. For passive particles and rather uniform nutrient profiles at larger $D_N$, fluctuations in the small initial bacterial distribution amplify in the transient phase because cells can form clusters throughout the entire system. This results in an asymmetric bacterial distribution during the transient phase, which dissolves when approaching the steady state. In a peaked nutrient distribution occurring at small $D_N$, bacterial growth is limited to a small region during the entire transient phase and the bacterial distribution looks rather symmetric around the nutrient source.

Interestingly, in non-motile systems the bacterial swarm and the nutrient roughly occupy the same area, which influences population dynamics. If the nutrient and thereby the bacterial cluster are strongly localized, the population number quickly approaches the steady state. We propose this is due to the localized nutrient almost entirely covered by bacteria, which results in a strong coupling. At increased nutrient diffusion, where the profile is more uniform, the nutrient strongly overshoots and is followed by a population overshoot. Here, nutrient and bacteria cover a larger area and the coupling is weaker. 

In contrast, highly active particles cover the entire system area even for peaked nutrient distributions and the trend reverses. For strongly localized nutrient at small diffusion constant the bacterial population strongly overshoots because a large fraction of the bacteria do not interact with the nutrient, which means weak coupling. If the nutrient profile becomes flatter, the coupling is stronger since a larger fraction of the bacteria can interact with the nutrient.

The observed difference in the dynamics can be rationalized by a uniform population model, where we introduce separate areas of the bacterial ($A_N$) and nutrient ($A_S$) distributions. Using $A_N=A_S$ for the passive case and $A_N = A_\mathrm{box}$ for the case of high activity, the population dynamics can be well fitted with a single parameter $A_S$. The relation between diffusion coefficient and effective nutrient area is monotonous in both cases. In other words, a more heterogeneous nutrient distribution results in a smaller effective system size. For the intermediate case of weakly active particles, the uniform model has limited applicability in the case of slow nutrient diffusion, where both limiting cases of zero and high activity cannot be applied.

We demonstrated that our minimalistic model allows us to build a bridge between analytic population models and particle-based simulations and to highlight differences between dividing active and passive particles. The model makes general statements that serve as a guideline for future theoretical and experimental investigations. Our insights into the role of nutrient heterogeneity can help to understand how an environment influences the individual reproductive success~\cite{tuljapurkar2020skewed,crow1989some,snyder2018pluck,caswell2011beyond}, the spread of neutral mutations~\cite{champagnat2012birth,bornholdt1998neutral}, and the population dynamics in changing environments~\cite{humston2004behavioral}. To make quantitative predictions, our model can be extended to include processes such as chemotaxis~\cite{engelmann1881neue,marsden2014chemotactic,saragosti2011directional,seyrich2019traveling,sturmer2019chemotaxis}, density/pressure-dependent growth rates~\cite{shraiman2005mechanical,basan2009homeostatic,ranft2010fluidization}, cell elongation~\cite{wittmann2022mechano,you2018geometry}, polydispersity~\cite{siebers2023exploiting,gompper20202020}, and more complex setup geometries.

\section*{Acknowledgments}
We thank Josua Grawitter, Arne W Zantop and Arnold JTM Mathijssen for interesting discussions and TU Berlin and the CRC 910 for financial support. TW also acknowledges support through a Deutschlandstipendium.

\section*{Conflict of interest}
The authors declare no competing interests.

\section*{References}

\bibliographystyle{iopart-num.bst}

\bibliography{Literature}

\newpage

\clearpage

\appendix

\section{Rescaled particle-based model}\label{sec:rescales}
The dynamics of the particle-based model (discussed in Sec.~\ref{sec:particle_based}) can be rescaled. We choose characteristic distance $\sigma$, time $\tau_R = D_R^{-1}$, energy $k_B T$, and nutrient amount $\gamma$. The rescaled variables are marked with a tilde: 
\begin{eqnarray*} 
    \mathrm{P} =    v  \sigma^{-1}\tau_R,\, \tilde g_\mathrm{max} = \tau_R g_\mathrm{max} ,\, \tilde K_s = \sigma^2 \gamma^{-1}K_s, \tilde d = \tau_R d,\,  \tilde D_N = \sigma^{-2}\tau_R D_N,\, \\ \tilde R_0 = \tau\gamma^{-1} R_0 ,\,
    \tilde k = k_B^{-1}T^{-1} \sigma^2 k \text{, and } \tilde v_d = \tau_R\sigma^{-1} v_d\,.
\end{eqnarray*}
The bacterial motion described in Eq.~(\ref{eq:dyn_position}) rescale to
\begin{eqnarray}
    d \tilde{\mathbf{r}}_i &= d\tilde{t} \left[P \mathbf{u}_i + \frac{1}{3} \sum_{i\neq j} \tilde{\mathbf{F}}_{ij}  \right]   
    + \sqrt{2D_T/D_R d\tilde{t} \bm{\xi}_i}\,,\\
    \varphi_i &= \sqrt{2d\tilde{t}} \eta_i \,.
\end{eqnarray}
with persistence number $\mathrm{P}$ and rescaled force $\tilde{\mathbf{F}}_{ij}= \tilde{k} \theta(|\tilde{\mathbf{r}}_{ij}|-1) (|\tilde{\mathbf{r}}_{ij}|-1) \hat{\tilde{\mathbf{r}}}$. The nutrient dynamics in Eq.~(\ref{eq:nutrient_individual}) rescales to:
\begin{eqnarray} 
        \frac{\partial \tilde{s}(\tilde{\mathbf{r}},\tilde{t})}{\partial \tilde{t}} = 
        \tilde{D}_N \tilde{\nabla}^2 \tilde{s}(\tilde{\mathbf{r}},\tilde{t}) + \tilde{R}_0\cdot\tilde\delta(\tilde{\mathbf{r}}-\tilde{\mathbf{r}}_s)
        -\sum\limits_{i=0}^{N}\tilde\delta(\tilde{\mathbf{r}}-\tilde{\mathbf{r}}_i(\tilde{t})) \tilde{g}(\tilde{s}(\tilde{\mathbf{r}}_i,\tilde{t}))\,.
\end{eqnarray}
with rescaled growth rate $\tilde g(\tilde s) = \tilde g_\mathrm{max} \tilde s /(\tilde K_s + \tilde s)$. The division rule simply rescales to \begin{center}\parbox{\dimexpr\linewidth-10\fboxsep-10\fboxrule\relax}{\textit{In a time interval $d\tilde t$, each particle has the probabilities of $d\tilde t \cdot \tilde{g}(\tilde{s}(\tilde{\mathbf{r}}_i,\tilde{t}))$ to divide and $d\tilde{t} \cdot \tilde{d}$ to die.}}\end{center}

\section{Biological Parameters}\label{sec:parameters}
The biological model of this paper is \textit{Escherichia coli}. The system is held at their optimum growth temperature $T_0 = \SI{41}{^\circ C}$~\cite{TEMEPRATURE_GROWTH_MODEL2}. At this temperature, water has a viscosity of $\eta_0 = \SI{0.6535}{mPas}$~\cite{VISCOSITY_OF_WATER}. \textit{E. coli} are  $2-5\,\si{\micro m}$ long and have a diameter of $0.8-1\,\si{\micro m}$~\cite{BACTERIAL_GROWTH}. They are approximated as discs of diameter $\sigma = 2\,\si{\micro m}$. The resulting orientation correlation time is $\tau_R = D_R^{-1} = \pi \eta \sigma^3/k_B T_0 = \SI{3.787}{s}$. The parameter $\gamma =\SI{2.99e-14}{mol}$ describes the characteristic amount of glucose required to make a new cell~\cite{BACTERIAL_GROWTH}. The diffusion coefficient of glucose $D_{N}(T=\SI{41}{^\circ C}, \eta =  \SI{0.6535}{mPas})=\SI{0.99}{\micro m^2ms^{-1}}=\SI{935}{\sigma^2\tau_R^{-1}}$ can be calculated using the linear scaling of the diffusion coefficient with $T$ and $\eta^{-1}$ and the reference value $D_{N}(T'=\SI{15}{^\circ C}, \eta' = \SI{1.1382}{mPas}) =\SI{0.52}{\micro m^2ms^{-1}}$~\cite{GLUCOSE_DIFFUSION_WATER,VISCOSITY_OF_WATER}. For \emph{E.\ coli} under glucose starvation, the parameters of the Monod function Eq.~(\ref{eq:monod}) are characteristic volume concentration $\mathcal{K}_s=\SI{22}{mol m^{-3}} = \SI{5.89e-3}{\gamma \sigma^{-3}}$~\cite{monod1949growth} and maximum growth rate $g_\mathrm{max}= \SI{2}{h^{-1}} = \SI{2.10e-3}{\tau_R^{-1}}$~\cite{BACTERIAL_GROWTH}. The starvation rate $d =  \SI{0.633}{day^{-1}} = \SI{2.77e-5}{\tau_R^{-1}} $ is extracted from a diagram of the reduction in bacterial culturability in~\cite{Starvation_Temperature}, for a more precise value further measurements are necessary.

\end{document}